\documentstyle[12pt]{article} 
 
\title{Quantum Zeno effect\\ in the decay onto an unstable level\thanks{Published in Phys. Lett. A~257, 227-231 (1999)}} 
\author{Michael B. Mensky\\ 
{\small P.N.Lebedev Physical Institute, 117924 Moscow, Russia}} 
\date{} 
\newcommand{\eq}[1]{(\ref{#1})} 
\newcommand{\eqs}[2]{(\ref{#1},~\ref{#2})} 
\newcommand{\Eq}{Eq.~\eq} 
\newcommand{\Eqs}{Eqs.~\eqs} 
\newcommand{\Sect}[1]{Sect.~\ref{#1}} 
\newcommand{\be}{\begin{equation}} 
\newcommand{\ee}{\end{equation}} 
\newcommand{\ba}{\begin{eqnarray}} 
\newcommand{\ea}{\end{eqnarray}} 
\newcommand{\ra}{\rangle} 
\newcommand{\la}{\langle} 
 
\newcommand{\g}{{\gamma}} 
\newcommand{\G}{{\Gamma}} 
\begin{document} 
\maketitle 
 
\begin{abstract} 
Under certain assumptions it is shown that the decay of level 2 of a three-level 
system onto level 1 is slowed down because of the further decay of level 1 onto  
level 0. It is argued that this phenomenon may be interpreted as  
a consequence of the quantum Zeno effect. The reason why this may be  
possible is that the second decay (or accompanying photon radiation)  
may be considered as a sign of the transition $2\rightarrow 1$ so that  
during the first transition the system is under continuous observation. 
\end{abstract} 
 
\section{Introduction}\label{intro} 
 
The quantum Zeno effect (paradox) \cite{Zeno-orig} is the name for the  
phenomenon of freezing (or slowing down) the evolution of a  
continuously observed quantum system. Originally the effect has been  
discussed in the case of a spontaneously decaying system, and  
preventing (or slowing down) of the decay has been predicted. Later on  
it was argued \cite{noZenoDecay} that the Zeno effect cannot arise in  
the spontaneous decay. Instead, the Zeno effect has been thoroughly  
investigated and finally experimentally proved \cite{Itano90-Zeno} in  
a repeatedly measured two-level system otherwise undergoing Rabi  
oscillations. The possibility of the Zeno effect in the initial  
non-exponential stage of a spontaneous decay is yet under discussion  
\cite{PlenioKnightThomps96decayZeno}. It was argued \cite{Panov96zeno}  
that the Zeno effect is observed in the real radioactive decay. A  
review on the subject can be found in \cite{HomeWhitaker97revZeno}. 
 
We shall present below a simple model predicting slowing down of a  
spontaneous decay in the case if the final (after this decay) state of  
the system is also unstable. This situation could in principle be  
interpreted as the Zeno effect in the continuously observed  
spontaneous decay, the second decay serving as a mechanism for the  
observation of the first one. 
 
We shall consider a 3-level system with level 2 spontaneously decaying 
onto level 1 and level 1 spontaneously decaying onto level 0. If the 
system is originally on level 2, then the decay  $1\rightarrow 0$ 
(practically, observation of a photon radiated simultaneously with  
this decay) is a sign that the system has already arrived at level 1  
and therefore that the transition  $2\rightarrow 1$ occurred. Vice  
versa, the absence of the decay  $1\rightarrow 0$ means that the  
system is yet at level 2. Thus, the very possibility of the decay   
$1\rightarrow 0$ means that the system prepared originally at level 2  
is under permanent observation (measurement). Then, as a result of the  
Zeno effect, the system must be frozen at level 2 or at least the  
decay of this level must be essentially slowed down. 
 
In \Sect{SectDecayDecay} we shall confirm by a direct  
quantum-mechanical calculation that this is the case: the decay   
$2\rightarrow 1$ is slowed down if level 1 is unstable, the greater  
instability of level 1, the less the rate of the decay $2\rightarrow  
1$. In \Sect{SectConclus} we shall return to the question whether this  
phenomenon can be interpreted as a result of the Zeno effect.  
 
To make the calculation more clear, we shall consider in  
\Sect{SectDecay} the decay onto a stable level and then in  
\Sect{SectDecayDecay} the model will be generalized to the case of  
interest. 
 
\section{The decay onto a stable level}\label{SectDecay} 
 
As the preliminary step, let us consider, by the method given in 
\cite{Peres-bk93}, a model of the decay $2\rightarrow 1$ onto a stable 
level 1. 
 
Let $H_0$ be a Hamiltonian of a multilevel system (atom) including 
also a continuous spectrum. The latter may originate from the 
interaction between the atom and the electromagnetic field (photons) 
which could be absorbed or radiated simultaneously with transitions of 
the atom. The nature of the continuous spectrum may be arbitrary, but 
for concreteness we shall speak of photons. The total Hamiltonian 
$H=H_0+V$ will contain also a potential $V$ leading to the transition 
between levels accompanying by the photon number change. 
 
Denote the state of the atom on level 2 by $|2\ra$. Suppose that there 
is no photons (more generally, no contribution from the continuous 
spectrum) in the state $|2\ra$. We wish to describe the decay of this 
state to the state $|1E\ra$ in which the atom is on level 1 and there 
are also some photons, so that the total energy of the atom and the 
electromagnetic field is $E$. For simplicity we shall assume that the 
only non-zero matrix elements of the potential $V$ are $\la 
1E|V|2\ra=\overline{\la 2|V|1E\ra}$. 
 
To describe the transition $|2\ra\rightarrow |1E\ra$, consider the 
general state of the system in the form 
\be 
|\psi\ra=a_2(t)\, |2\ra \, e^{-iE_2 t} + 
        \sum_E a_{1E}(t)\, |1E\ra\, e^{-iEt} 
\label{gen-state}\ee 
where the natural units ($\hbar=1$) are used and the integration in 
energy is denoted as a sum. To return to the usual units, we have to 
replace $t$ by $t/\hbar$. To return to the genuinely continuous 
spectrum, we have to replace a sum over $E$ by integration over $E$ 
with the weight $\rho_1(E)$ presenting the local density of states 
$|1E\ra$. 
 
Substituting this form for the state in the Schr\"odinger equation, we 
have the following equations for the coefficients $a_2$, $a_{1E}$: 
\ba 
\dot a_2\, e^{-iE_2t}&=&-i\sum_E \la 2|V|1E\ra\,a_{1E}\, e^{-iEt}, 
\label{coef-eq1}\\ 
\dot a_{1E}\, e^{-iEt}&=& -i\; \la 1E|V|2\ra\,a_2\, e^{-iE_2t}. 
\label{coef-eq2}\ea 
To solve these equations, let us accept the anzatz $a_2(t)=\exp(-\g_2 
t)$ corresponding to the exponential law of the decay of level 2 (this 
law is valid for not too small times). Then \Eq{coef-eq1} will take 
the form 
\be 
i\sum_E a_{1E}\la 2|V|1E\ra\,e^{-i(E-E_2)t}=\g_2\, e^{-\g_2 t} 
\label{coef-eq1a}\ee 
while \Eq{coef-eq2} may be explicitly solved to give 
\be 
a_{1E}(t)=\frac{\la 1E|V|2\ra}{E-E_2+i\g_2} 
\left[ 1-e^{i(E-E_2+i\g_2)t} 
\right]. 
\label{sol1}\ee 
The initial condition $a_{1E}(0)=0$ is used to describe the system 
being initially on level 2. 
 
Now we have to substitute the expression \eq{sol1} for the function 
$a_{1E}(t)$ in \Eq{coef-eq1a}. Evaluating the sum (integral) on 
energies in \Eq{coef-eq1a}, we shall assume that the weight function 
$\rho_1(E)$ and the matrix element $\la 1E|V|2\ra$ are slow functions 
of energy and can be replaced by the constants equal to the values of 
these functions at $E=E_2$ (the energy of level 2, the point where the 
denominator in \Eq{sol1} has minimum). Under this assumption the 
energy integral can be evaluated. \Eq{coef-eq1a} may be shown to be 
satisfied provided that 
\be 
\g_2=\pi\, \rho_1(E_2)\, |\la1E_2|V|2\ra|^2. 
\label{g2}\ee 
This is nothing else than the ``Fermi's golden rule" for the decay of 
an unstable level. 
 
\section{The decay onto a decaying level}\label{SectDecayDecay} 
 
Let us apply an analogous consideration to the three-level system of 
interest: level 2 may decay to level 1, and level 1 in turn may 
decay to level 0. The general state of the system (again 
containing a continuous spectrum, photons) may be presented in the 
form 
\be 
|\psi\ra=a_2(t)\, |2\ra \, e^{-iE_2 t} + 
  \sum_E (a_{0E}(t)\, |0E\ra+\sum_E a_{1E}(t)\, |1E\ra)\, e^{-iEt}. 
\label{gen-state2}\ee 
Here $|1E\ra$ denotes the state of the atom at level 1 and the general 
energy of the system (atom plus photons) $E$, $|0E\ra$ is an analogous 
state but with the atom at level 0. The sums over energies will be 
later replaced by the integrals with the corresponding weights: 
$\rho_1(E)$ for the states $|1E\ra$ and $\rho_0(E)$ for $|0E\ra$. The 
Hamiltonian of the system will be taken in the form $H=H_0+V$ with the 
following non-zero matrix elements of $V$: $\la 1E|V|2\ra$ and $\la 
0E|V|1E'\ra$. Then the Schr\"odinger equation gives the following 
equations for the coefficients: 
\ba 
\dot a_2&=&-i\sum_E \la 2|V|1E\ra\, a_{1E}\, e^{-i(E-E_2)t} 
\label{coef-eq32}\\ 
\dot a_{1E}&=& 
   -i\la 1E|V|2\ra\, a_2\, e^{-i(E_2-E)t}- 
   i\sum_{E'}\la 1E|V|0E'\ra\, a_{0E'}\, e^{-i(E'-E)t}\label{coef-eq31}\\ 
\dot a_{0E}&=&-i\sum_{E'}\la 0E|V|1E'\ra\, a_{1E'}\,e^{-i(E'-E)t}. 
\label{coef-eq30}\ea 
 
To solve this set of equations, we shall present them in the vector 
form 
\ba 
\dot a_2&=&-i\,V_{21}\, a_1 \label{coef-vect2}\\ 
\dot a_1&=& 
   -i\,V_{12}\, a_2 - i\,V_{10}\, a_0 \label{coef-vect1}\\ 
\dot a_0&=&-i\,V_{01}\, a_1 
\label{coef-vect0}\ea 
where the following vectors and matrices are introduced: 
\ba 
(a_1)_E=a_{1E}, \quad (a_0)_E=a_{0E}, \nonumber\\ 
(V_{21})_E=\la 2|V|1E\ra\,e^{-i(E-E_2)t}, \quad 
(V_{12})_E=\la 1E|V|2\ra\,e^{-i(E_2-E)t}, \nonumber\\ 
(V_{10})_{EE'}=\la 1E|V|0E'\ra\,e^{-i(E'-E)t}, \quad 
(V_{01})_{EE'}=\la 0E|V|1E'\ra\,e^{-i(E'-E)t}. 
\label{vectors}\ea 
 
Let us introduce also the integral operations acting on the 
time-dependent vectors: 
\be 
I_{kl}\,a=-i\int_0^tV_{kl}\,a\, dt, \quad J=I_{10}I_{01}. 
\label{Integrals}\ee 
Then \Eqs{coef-vect1}{coef-vect0} may be replaced by the integral 
equations 
\be 
a_1=I_{12}\,a_2+I_{10}\, a_0, \quad a_0=I_{01}\, a_1 
\label{Int-eq}\ee 
having the solution 
\be 
a_1=(1-J)^{-1}I_{12}\, a_2=\sum_{n=0}^\infty J^n \,I_{12}\, a_2. 
\label{sol-int}\ee 
Making use of the anzatz $a_2=e^{-\G_2t}$ in the right-hand side of 
this equation and substituting the resulting expression for $a_1$ in 
\Eq{coef-vect2}, we have the following equation for $\G_2$: 
\be 
-i\sum_{n=0}^\infty V_{21}\, J^n \,I_{12}\, e^{-\G_2t}=-\G_2\, e^{-\G_2t}. 
\label{EqG2}\ee 
 
We can evaluate each term in the sum. For calculating sums (integrals) 
over energies, we shall use the same approximation as in the preceding 
section considering all matrix elements of $V$ and the weight 
functions $\rho_1(E)$ for states $|1E\ra$ and $\rho_0(E)$ for states 
$|0E\ra$ slow functions of energies. Then each term in the left-hand  
side of \Eq{EqG2} can be evaluated.  
 
It turns out that the terms corresponding to the given $n$ differ from  
the term corresponding to $n-1$ only by the numerical factor  
$(-N)$ where  
\be 
N=\pi^2\rho_0(E_2)|\la 0E_2|V|1E_2\ra|^2 \rho_1(E_2). 
\label{N}\ee 
This gives  
\be 
\G_2=\frac{\g_2}{1+N} 
\label{G2}\ee 
where $\g_2$ is defined by \Eq{g2}.  
\Eq{G2} is proved in the assumption that $N<1$, however this does not  
exclude that it may be valid also in a wider region. The assumptions  
about the behavior of matrix elements of $V$ and functions $\rho_1$  
and $\rho_2$ taken above are essential.  
 
The formula \eq{G2} leads to the main conclusion. The entity $N$ in 
its denominator is proportional to the rate of the decay of level 1 in 
the situation when the system starts in the state $|1E_2\ra$. In other 
words, $N$ is a measure of instability of level 1 (under the condition  
that there are also photons so that the total energy of the system is  
$E_2$). We see therefore that the rate of the decay of level 2  
decreases because of instability of the target level 1. The more  
instability of level 1, the less the rate of the decay $2\rightarrow  
1$. 
 
The last claim may be made more concrete if we (roughly) estimate the  
rate $\G_1$ of the decay of level 1. It depends on the energy band of  
the decaying states $|1E\ra$. Since these states result in the decay  
of the state $|2\ra$, the energy $E$ should be of the order of $E_2$  
and the width $\Delta E$ of the energy band is of the order of $\G_2$.  
The rate of the decay of the level 1 may be obtained (as a rough  
estimate) by multiplication of the number \eq{N} by $\Delta E$ giving  
$\G_1\sim\G_2\,N$. According to \Eq{G2}, the slowing down of the decay  
$2\rightarrow 1$ is essential if $\G_1$ is larger than or of the order  
of $\G_2$. 
 
\section{Discussion}\label{SectConclus} 
 
We showed, under certain assumptions, that the rate of the decay is 
slowed down by instability of the target level. It has been argued in 
Introduction that this may be interpreted as a consequence of the 
quantum Zeno effect. However, this must be compared with the 
arguments against the Zeno effect in spontaneous decays. 
 
Some authors argued \cite{noZenoDecay} that the quantum Zeno effect 
is impossible in spontaneous decay because of its exponential law 
(contrary to the quadratic small-time asymptotic of Rabi oscillations). 
One more doubt may be based on the  following. A spontaneously 
decaying atom located in plasma is subject  to repeated scattering of 
electrons on the atom. These events of  scattering may be thought of as 
repeated measurements of the atom discriminating its levels. In this 
situation the Zeno effect, if existing, could slow down the decay. In reality 
the rate of the decay changes insignificantly \cite{VainshSobelmanYuk79bk} 
due to scattering of electrons. 
 
This gives an additional argument against the Zeno effect in spontaneous 
decay. The phenomenon discussed in the preceding section may then be 
interpreted as a Zeno-like but not genuinely Zeno effect. 

In our opinion, the conclusion should not be as radical as this. Instead, 
one may consider the possibility of the Zeno effect for different types of 
continuous measurements. In the situation considered in the present 
paper the transition $1\rightarrow 0$ is an evidence that the system has 
already arrived at level 1, but after this evidence has been obtained the 
system is no more at level 1. If we consider a 2-level system with both 
levels 2 and 1 as our measured system, then the measurement leads to 
destruction of this system. On the contrary, scattering electrons gives 
information about the level the atom is on and leaves it at the same level. 
The measurements described by von Neumann's projections 
(discussed in most papers on the Zeno effect) act analogously. 
 
Therefore, the results obtained in the present may point out that  
1)~the quantum Zeno effect does not arise in a spontaneous decay if  
the measurement is ``minimally disturbing" (described by projectors),  
but 2)~the effect takes place if the measurement is ``destructive"  
i.e. leads to the disappearance of the measured state.  

Some other remarks must be added. The conclusion about slowing down 
the decay $2\rightarrow 1$ due to the decay $1\rightarrow 0$ has been 
proved above under certain assumptions. The most important of them are 
that the phase volumes of the transitions may be correctly accounted by 
the weights $\rho_1$, $\rho_0$ and these weights are slow 
functions of energy $E$. Under other conditions the conclusion about 
slowing down of the decay would be different. This may be considered as 
one more argument against the Zeno interpretation of slowing down, but 
instead this may point out on necessity of more accurate treatment of the 
concept of ``observation".  

Observation of the decay is characterized by the time of observation and 
energy of the decay products. Simple statements that the decay ``is 
observed" or ``is not observed" are hardly quite adequate. One needs 
quantitative characteristics of the observations. The degree of slowing down 
the decay must depend on these characteristics. The assumption accepted 
in the present paper that the functions $\rho_1(E)$, $\rho_0(E)$ are slowly 
varying, means that the products of the decays are efficiently observed in a 
wide energy band. In this condition the Zeno effect may be expected. If the 
functions $\rho_1(E)$, $\rho_0(E)$ have the shape of narrow peaks, the 
Zeno effect may be absent because of inefficient observation. 

It may be remarked in this connection that the very existence of the decay 
products may naively be considered as an evidence of the decay. If one  
accept this point of view, he is forced to conclude that the decay is always 
under continuous observation and therefore is always subject to the Zeno 
effect. This is however invalid (see the paper of A.Peres in \cite{Zeno-orig}) 
because the decay products have to be considered as a part of the system 
necessary for the description of the decay itself. 

The secondary decay in a three-level system analyzed in the present paper 
may be in most cases considered as external to the primary decay. Hence, 
this secondary decay may be treated as an observation and must lead to 
slowing down of the primary decay. Even in this case much more accurate 
and detailed analysis is necessary to have complete and reliable description 
of the Zeno effect. This analysis has to include all temporal and energy 
characteristics of the process. As a limiting case, it cannot be excluded that 
in some conditions (for certain characteristics of the system and its 
environment) the secondary decay cannot be considered as being external 
in respect to the primary decay. 

Summing up, we suggest that the complete analysis of the Zeno effect in decay 
requires more detailed definition of the concept of observation and of the Zeno 
effect itself. The results of the present paper show that the development of this 
sort must be fruitful. 
 
\vskip 0.5cm 
\centerline{\bf ACKNOWLEDGEMENT} 

The author acknowledges the fruitful discussions with V.~Namiot and A.~Panov 
concerning interpretation of the results. The work was supported in part by the 
Russian Foundation of Basic Research, grant 98-01-00161.

\end{document}